# The stability of 12-fold symmetry soft-matter quasicrystals

Tian-You Fan[1]* and Zhi-Yi Tang[2]
1 School of Physics, Beijing Institute of Technology, Beijing 100081, China
2 School of Computer Science and Technology, Beijing Institute of Technology, Beijing 100081, China
*Corresponding author, email:tyfan2013@163.com

**Abstract** This letter presents a study on the stability of the 12-fold symmetry soft-matter quasicrystals from the angle of thermodynamics combining dynamics of the matter. The results are quantitative, which depend upon only the material constants of the novel phase and very simple and intuitive, these material constants can be measured by experiments.

## 1. Introduction

The soft-matter quasicrystals have been observed over 15 years in liquid crystals, polymers, colloids, nanoparticles and surfactants and so on [1-12], in which the most frequently observed is 12-fold symmetry ones, which become the most important soft-matter quasicrystals, although later the 18-fold symmetry soft-matter quasicrystals are also found but only in colloids. Based on the symmetry theory, the 5-,8- and 10-fold symmetry structures are similar to the 12-fold symmetry ones, which will be observed in the near future, we call them the first kind of soft-matter quasicrystals. From point of view of symmetry, the 7-, 9- and 14-fold symmetry quasicrystals are similar to those of 18-fold symmetry ones, which will be found in the near future, they can be identified as the second kind of soft-matter quasicrystals. The discovery of soft-matter quasicrystals is an important event in 21$^{th}$ century chemistry. The soft-matter quasicrystals are formed through self-assembly of spherical building blocks by supramolecules, compounds and block copolymers and so on, which is associated with chemical process and is quite different from that of solid quasicrystals, i.e., the binary and ternary metallic alloy quasicrystals which are formed under rapid cooling. These two thermodynamic environments are completely different to each other. The new structure presents both natures of soft matter and quasicrystals. Soft matter is an intermediate phase between ideal solid and simple fluid, which exhibits fluidity as well as complexity as pointed out by de Gennes [13], while for quasicrystals the symmetry is very important as they are highly ordered phase. We can say the soft-matter quasicrystals are complex fluid with quasiperiodic symmetry. Refs [14-19] reviewed soft-matter quasicrystals from different angles on their formation mechanism, structure stability, thermodynamics and the correlation between Frank-Kasper phase and quasicrystals etc., in which stability of the soft-matter quasicrystals discussed by Lifshitz and Diamant [19] in 2007 soon after the discovery of the new phase. They pointed out that because the quite difference formation mechanism of soft-matter quasicrystals with the solid ones, so contrary to the latter, the source of stability soft-matter quasicrystals remains a question of great debate to this day. Since then up to now, the topic on stability of soft-matter quasicrystals are interested by a quite lot of researchers [20-24]. They gave a different treatment, this shows the problem is complicated. Lifshitz and Diamant suggested to study the problem from the effective free energy that proposed by Lifshitz and Petrich[25], this energy is connected with mass density $\rho$, the governing equation describing the time-space evolution of $\rho$ is given by Lifshitzand Diamant, they and co-workers did many analyses on the stability of 12-fold symmetry soft-matter. The other discussions on the stability in Refs [19-24] etc concern thetwo natural length scales and two wave numbers, which are heritage and development of the Lifshitz's pioneering work and promote the study.

Different from the effective free energy approach in studying the stability suggested by Lifshitz and other physicists and mathematicians, we would like to give a probe by a combination between dynamics and thermodynamics. Our group works the generalized dynamics of the soft-matter quasicrystals over the years [26-29], we find some results of the dynamics might be used by the thermodynamics, this simplifies the discussion on the stability and easily to find some quantitative results.For this purpose we suggest an extended free energy of the quasicrystal system in soft matter which is the key of the following discussion.

## 2. Extended free energy of the quasicrystal system in soft matter

The soft-matter quasicrystalis a complex viscous and compressible fluid with quasiperiodic symmetry, consists of elementary excitations phonon displacement field $\mathbf{u}(u_x, u_y, u_z)$, phason displacement field $\mathbf{w}(w_x, w_y, w_z)$ and fluid phonon velocity field $\mathbf{V}(V_x, V_y, V_z)$ and

$$\varepsilon_{ij} = \frac{1}{2}\left(\frac{\partial u_i}{\partial x_j} + \frac{\partial u_j}{\partial x_i}\right), w_{ij} = \frac{\partial w_i}{\partial x_j}, \dot{\xi}_{ij} = \frac{1}{2}\left(\frac{\partial V_i}{\partial x_j} + \frac{\partial V_j}{\partial x_i}\right) \quad (1)$$

aretensors of phonon strain, phason strain and fluid deformation rate, respectively [26-29], then we define the extended inner energy density

$$U_{ex} = \frac{1}{2} A \left(\frac{\delta\rho}{\rho_0}\right)^2 + B\left(\frac{\delta\rho}{\rho_0}\right)\nabla\cdot\mathbf{u} + C\left(\frac{\delta\rho}{\rho_0}\right)\nabla\cdot\mathbf{w} + U_{el} \quad (2)$$

where the first term denotes energy density due to mass density variation, the quantity $\frac{\delta\rho}{\rho_0}$ describes the variation of the matter density, in which $\delta\rho = \rho - \rho_0$ and $\rho_0$ the initial mass density, according to our computation in the cases of transient response and flow past obstacleof soft-matter quasicrystals [29], $\frac{\delta\rho}{\rho_0} = 10^{-4} \sim 10^{-3}$ for soft-matter quasicrystals, which describes the fluid effect of the matter and is greater 10 order of magnitude than that of solid quasicrystals (in this sense we can consider for the solid quasicrystals the effect of $\frac{\delta\rho}{\rho_0}$ is very weak); the second term in (2) is one by mass density variation coupling phonons; the third represents that of mass density variation coupling phasons, and $A$, $B$ and $C$ the corresponding material constants, respectively. According to [30] $C$ should be zero. Based on our computations due to extremely smaller of value of $\nabla\cdot\mathbf{w}$ than that of $\nabla\cdot\mathbf{u}$, the term

$C\left(\dfrac{\delta\rho}{\rho_0}\right)\nabla\cdot\mathbf{w}$ can be omitted hereafter. We should point out energy density of mass density variation, mass density variation coupling phonons and phasonsare suggested by Lubensky et at [30]for the first time from the hydrodynamics of solid qyasicrystals, and the present generalized dynamics of soft-matter quasicrystals is developed drawn from the theory of Lubensky et al, in particular the Hamiltonian

$$\left.\begin{aligned}H &= H[\Psi(\mathbf{r},t)]\\ &= \int \dfrac{\mathbf{g}^2}{2\rho}d^d\mathbf{r} + \int\left[\dfrac{1}{2}A\left(\dfrac{\delta\rho}{\rho_0}\right)^2 + B\left(\dfrac{\delta\rho}{\rho_0}\right)\nabla\cdot\mathbf{u}\right]d^d\mathbf{r} + F_{el}\\ &= H_{kin} + H_\rho + F_{el}\\ F_{el} &= F_u + F_w + F_{uw}, \qquad \mathbf{g} = \rho\mathbf{V}\end{aligned}\right\} \quad (3)$$

is drawn from Lubensky et al [30], in which $H_{kin}$ denotes the kinetic energy, $H_\rho$ the energy due to the variation of mass density, $F_{el}$ the elastic deformation energy consisting of contributed from phonons, phasons and phonon-phason coupling, respectively, the detailed definition will be given by (4)-(6) in the following. However, the dynamics equations of soft-matter quasicrystals are different from those of hydrodynamics of solid quasicrystals, and some results of Lubensky equations originated from [30] are not the same in Refs [26-29], for example, the equation (A2) in [30]

$$\dfrac{\delta\rho}{\rho_0} = -\dfrac{B}{A}\nabla\cdot\mathbf{u}$$

does not hold for the dynamics of soft-matter quasicrystals etc.

In equation (2)

$$U_{el} = U_u + U_w + U_{uw} \quad (4)$$

represents the elastic free energy density come from phonons, phasons and phonon-phason coupling such as

$$\left.\begin{aligned}U_u &= \dfrac{1}{2}C_{ijkl}\varepsilon_{ij}\varepsilon_{kl}\\ U_w &= \dfrac{1}{2}K_{ijkl}w_{ij}w_{kl}\\ U_{uw} &= R_{ijkl}\varepsilon_{ij}w_{kl} + R_{klij}w_{ij}\varepsilon_{kl}\end{aligned}\right\} \quad (5)$$

according to the constitutive law of soft-matter quasicrystals [26-29], i.e., we have the constitutive law for the matter

$$\left.\begin{aligned}\sigma_{ij} &= C_{ijkl}\varepsilon_{kl} + R_{ijkl}w_{kl}\\ H_{ij} &= K_{ijkl}w_{kl} + R_{klij}\varepsilon_{kl}\\ p_{ij} &= -p\delta_{ij} + \sigma'_{ij} = -p\delta_{ij} + \eta_{ijkl}\dot{\xi}_{kl}\end{aligned}\right\} \quad (6)$$

where $C_{ijkl}$ denotes the phonon elastic constants, $K_{ijkl}$ the phason elastic constants, and $R_{ijkl}, R_{klij}$ the phonon-phason coupling elastic constants [26-29] respectively. The elastic energy in (2) is the integrations of the relevant density of (5) over the domain. According to the thermodynamics, the extended free energy density is defined by

$$F_{ex} = U_{ex} - TS \quad (7)$$

where $U_{ex}$ the extended inner energy density defined by (2), and the $T$ absolute temperature and $S$ the entropy, respectively.

**Lemma**

From (7) and (2),(4)-(6), we have

$$S = -\frac{\partial F_{ex}}{\partial T}, \sigma_{ij} = \frac{\partial F_{ex}}{\partial \varepsilon_{ij}}, H_{ij} = \frac{\partial F_{ex}}{\partial w_{ij}}, \delta^2 F_{ex} \geq 0 \quad (8)$$

in which the second and third terms are equivalent to the elastic constitutive law of the material for degrees of freedom of phonons and phasons, and the last one is the stability condition of the matter, respectively. Because the formula (2) introduced an extended inner energy density, the variation principle containing in equations (8) is an extended or generalized variation.

## 3. The positive definite nature of the rigidity matrix and the stability of the soft-matter quasicrystals with 12-fold symmetry

For point group $12\,mm$ of the 12-fold symmetry quasicrystals in soft matter the concrete constitutive law is as following

$$\left.\begin{aligned}
\sigma_{xx} &= C_{11}\varepsilon_{xx} + C_{12}\varepsilon_{yy} + C_{13}\varepsilon_{zz} \\
\sigma_{yy} &= C_{12}\varepsilon_{xx} + C_{11}\varepsilon_{yy} + C_{13}\varepsilon_{zz} \\
\sigma_{zz} &= C_{13}\varepsilon_{xx} + C_{13}\varepsilon_{yy} + C_{33}\varepsilon_{zz} \\
\sigma_{yz} &= \sigma_{zy} = 2C_{44}\varepsilon_{yz} \\
\sigma_{zx} &= \sigma_{xz} = 2C_{44}\varepsilon_{zx} \\
\sigma_{xy} &= \sigma_{yx} = 2C_{66}\varepsilon_{xy} \\
H_{xx} &= K_1 w_{xx} + K_2 w_{yy} \\
H_{yy} &= K_2 w_{xx} + K_1 w_{yy} \\
H_{yz} &= K_4 w_{yz} \\
H_{xy} &= (K_1 + K_2 + K_3) w_{xy} + K_3 w_{yx} \\
H_{xz} &= K_4 w_{xz} \\
H_{yx} &= K_3 w_{xy} + (K_1 + K_2 + K_3) w_{yx} \\
p_{xx} &= -p + 2\eta \dot{\xi}_{xx} - \frac{2}{3}\eta \dot{\xi}_{kk} \\
p_{yy} &= -p + 2\eta \dot{\xi}_{yy} - \frac{2}{3}\eta \dot{\xi}_{kk} \\
p_{zz} &= -p + 2\eta \dot{\xi}_{zz} - \frac{2}{3}\eta \dot{\xi}_{kk} \\
p_{yz} &= 2\eta \dot{\xi}_{yz} \\
p_{zx} &= 2\eta \dot{\xi}_{zx} \\
p_{xy} &= 2\eta \dot{\xi}_{xy}
\end{aligned}\right\} \quad (9)$$

in which
$C_{1111} = C_{11}, C_{1122} = C_{12}, C_{3333} = C_{33},$
$C_{1133} = C_{13}, C_{2323} = C_{44}, C_{1212} = C_{66},$
$(C_{11} - C_{12})/2 = C_{66},$
$K_{1111} = K_{2222} = K_{2121} = K_{1212} = K_1,$
$K_{1122} = K_{2211} = -K_{2112} = -K_{1221} = K_2, K_{1122} = K_{1221} = K_{2112} = K_3, K_{2323} = K_{1313} = K_4$

the phonon-phason coupling constants $R_{ijkl} = R_{klij} = 0$ due to decoupling between phonons and phasons, i.e.,

$$U_{uw} = R_{ijkl}\varepsilon_{ij}w_{kl} + R_{klij}w_{ij}\varepsilon_{kl} = 0$$

for this type of quasicrystals.

As well as the conventional innerenergy density, between stresstensor and strain tensor there is aelastic rigidity matrix, for the extended inner energy density (2) there is an extended rigidity matrix such as

$$M = \begin{pmatrix} A & B & B & B & 0 & 0 & 0 & 0 & 0 & 0 & 0 & 0 & 0 \\ B & C_{11} & C_{12} & C_{13} & 0 & 0 & 0 & 0 & 0 & 0 & 0 & 0 & 0 \\ B & C_{12} & C_{11} & C_{13} & 0 & 0 & 0 & 0 & 0 & 0 & 0 & 0 & 0 \\ B & C_{13} & C_{13} & C_{33} & 0 & 0 & 0 & 0 & 0 & 0 & 0 & 0 & 0 \\ 0 & 0 & 0 & 0 & 2C_{44} & 0 & 0 & 0 & 0 & 0 & 0 & 0 & 0 \\ 0 & 0 & 0 & 0 & 0 & 2C_{44} & 0 & 0 & 0 & 0 & 0 & 0 & 0 \\ 0 & 0 & 0 & 0 & 0 & 0 & C_{11}-C_{12} & 0 & 0 & 0 & 0 & 0 & 0 \\ 0 & 0 & 0 & 0 & 0 & 0 & 0 & K_1 & K_2 & 0 & 0 & 0 & 0 \\ 0 & 0 & 0 & 0 & 0 & 0 & 0 & K_2 & K_1 & 0 & 0 & 0 & 0 \\ 0 & 0 & 0 & 0 & 0 & 0 & 0 & 0 & 0 & K_4 & 0 & 0 & 0 \\ 0 & 0 & 0 & 0 & 0 & 0 & 0 & 0 & 0 & 0 & K_1+K_2+K_3 & 0 & K_3 \\ 0 & 0 & 0 & 0 & 0 & 0 & 0 & 0 & 0 & 0 & 0 & K_4 & 0 \\ 0 & 0 & 0 & 0 & 0 & 0 & 0 & 0 & 0 & 0 & K_3 & 0 & K_1+K_2+K_3 \end{pmatrix}$$

(10)

Due to the condition in (8)

$$\delta^2 F_{ex} \geq 0 \quad (11)$$

thisnon-negative condition of the second order variation of the extended inner energy density functionalrequires the extended rigidity matrix must be positive definite. We have the theorem for describing the stability of the soft-matter quasicrystals with 12-fold symmetry as follows:

## Theorem

Under the condition (2), the validity of variation (11) is equivalent to the positive definite nature of matrix (10) and leads to

$$\left.\begin{array}{l} A > 0, \ A(C_{11}C_{33} + C_{12}C_{33} - 2C_{13}^2) - B^2(C_{11} + C_{12} - 4C_{13} + 2C_{33}) > 0, \ C_{11} - C_{12} > 0, \\ C_{44} > 0, \ K_1 - K_2 > 0, \ K_1 + K_2 > 0, \ K_1 + K_2 + 2K_3 > 0, \ K_4 > 0 \end{array}\right\} (12)$$

The proof of the theorem is straightforward.

When the conditions (12) are satisfied, the soft-matter quasicrystals of 12-fold symmetry are stable. This stability takes into account of the effects of fluid, fluid coupling phonons, phonons and phasons of soft-matter quasicrystals of 12-fold symmetry, more exactly speaking the stability depends upon the material constants concerning only on fluid, fluid coupling phonons, phonons and phasons of the soft-matter quasicrystals. These constants can be measured by experiments which are similar to that in crystallography [31] and solid quasicrystallography [32], and presentsimplicity and intuitive character of the complexityof stability of soft-matter quasicrystals. This shows the substantive nature of the stability of soft-matter quasicrystals. Substantively it explores the structure of the matter, because it comes from the constitutive law (9), which is the result of the symmetry of the structure---i.e., the result obtained by theory of group and group representation of the quasicrystals, refer to [26-29].

## 4. Comparison and examination

It is well-known that the 12-fold symmetry quasicrystals in soft matter belong to a type of two-dimensional quasicrystals, and in which the phonon field structure presents the character of the hexagonal crystals [26-29].

For the special case, as phason field is absence, i.e.,

$$K_1 = K_2 = K_3 = K_4 = 0 \quad (13)$$

and at the same time one takes $B=0$ and $A$ being any positive value, then (12) reduces to

$$C_{11} - C_{12} > 0, \quad C_{44} > 0, \quad C_{11}C_{33} + C_{12}C_{33} - 2C_{13}^2 > 0 \quad (14)$$

Here constant $A$ can be any positive value, and the value can be taken arbitrary small, due to $\left(\dfrac{\delta\rho}{\rho_0}\right)^2 \sim 10^{-6}$ (because $\dfrac{\delta\rho}{\rho_0} \sim 10^{-3}$, according to our computation for soft-matter quasicrystals), so $A\left(\dfrac{\delta\rho}{\rho_0}\right)^2$ is very small, this may be understood that if the fluid effect is very weak and no phason field, the stability condition (12) is reduced to that of hexagonal crystal system (14), the latter was derived by Cowley [31]. This, in one angle, is explored to examine the dynamics and thermodynamics of soft-matter quasicrystals being correct, but in another angle, it shows the constant $A$ could not be zero, this means the soft matter cannot be reduced to a solid phase from the angle of requirementof soft matter stability.

While for another case, i.e., in the solid quasicrystals of 12-fold symmetry, then (12)reduces to

$$\left.\begin{array}{l}C_{11} - C_{12} > 0, C_{44} > 0, C_{11}C_{33} + C_{12}C_{33} - 2C_{13}^2 > 0 \\ K_1 - K_2 > 0, K_1 + K_2 > 0, K_1 + K_2 + K_3 > 0, K_4 > 0\end{array}\right\} (15)$$

under condition $B=0$ and $A$ being any positive value, the inequality (15) is the stability condition of solid quasicrystals of 12-fold symmetry, which is derived by the authors and explores the correctness of the dynamics and thermodynamics of soft-matter quasicrystals once again, and shows again the soft matter cannot be reduced to a solid phase from the angle of requirement of soft matter stability.

The stability is connected with the positive definite nature of the mathematical structure of the matter dynamics, this is useful to the numerical solution (e.g. the finite element method), which will be discussed inouranother paper.

The stability is connected to the phase transition, this is more important problem. For crystals, Cowley [31] gave an analysis, and for quasicrystals, it requires to carry out to continue the probe.

## 5. Discussion and conclusion

By a complete different approach compared with Refs [19-24] this report discussed the stability of soft-matter quasicrystals, which directly is based on the thermodynamics with the help of generalized dynamics of the matter, and offered some quantitative and very simple results, the stability depends only upon the material constants, which can be measured by experiments. The correctness and precisionof the theoretical prediction isexamined by results of crystals and solid quasicrystals in qualitatively as well as quantitatively (refer to (14) and (15)).In the examination through the crystals and solid quasicrystals, we find that the constant $A$ can be arbitrary positive number but does not equal to zero, this shows the soft matter state cannot be reduced to any solid state phase, so the constant presents an important meaning for soft matter. The introducing of the constant $A$ was by Lubensky in hydrodynamics of solid quasicrystals [30] which is drawn by the first author of the paper to study the generalized dynamics of soft-matter quasicrystals over the years, due to the importance of the mass density in soft matter, the constant $A$ becomes more important in the matter in the study of dynamics as well as thermodynamics.

The discussion can also be given in accordance with the first law of thermodynamics, in this law we have

$$T = \frac{\partial U_{ex}}{\partial S}, \sigma_{ij} = \frac{\partial U_{ex}}{\partial \varepsilon_{ij}}, H_{ij} = \frac{\partial U_{ex}}{\partial w_{ij}}, \delta^2 U_{ex} \geq 0 \quad (16)$$

where $U_{ex}$ is defined by (2), which is a quadratic form, the condition $\delta^2 U_{ex} \geq 0$ requires the matrix (10) must be positive definite , so leads to the results (12), i.e., the theorem holds.

For possible soft-matter quasicrystals with 5-, 8- and 10-fold symmetry the stabilities are similar to that given by (11), we will report in other case.

The stability of the second kind of soft-matter quasicrystals can be seen the report given by [33].

**Acknowledgement** The work is supported by the National Natural Science Foundation of China through the grant 11272053. The first author of the article thanks Prof R Lifshitz of Tel Aviv University for presenting the electronic copy of Refs [19,20]. Zhi-Yi Tang is grateful to the support in part of the National Natural Science Foundation of China through the grant 11871098.

# References

[1]ZengX,Ungar G, Liu Y, Percec V, Dulcey A E and Hobbs J K ,Supramolecular dendritic liquid quasicrystals, *Nature*, 2004,**428**, 157-159.
[2] A. Takanoet al., A mesoscopic Archimedean tiling having a new complexity in an ABC star polymer, *J PolymSci Pol Phys*, 2005, **43**(18), 2427-2432.
[3]Hayashida K, Dotera T, Takano A, and Matsushita Y, Polymeric Quasicrystal: MesoscopicQuasicrystalline Tiling in A B C Star Polymers, *Phys. Rev.Lett.*,2007, **98**, 195502.
[4]Wasio, N. A. et al,Self-assembly of hydrogen-bonded two-dimensionalquasicrystals, *Nature*,2014,**507**, 86–89.
[5]Ye, X. et al, Quasicrystallinenanocrystalsuperlattice with partial matchingrules, *Nat. Mater.*,2016, **16**, 214–219.
[6]Talapin V D, Shevechenko E V, Bodnarchuk M I, Ye X C, Chen J and Murray C B, Quasicrystalline order in self-assembled binary nanoparticle superlattices,*Nature*, 2009,**461**, 964-967.
[7]Schulze M W, Lewis R M et al, Conformational Asymmetry and Quasicrystal Approximants in Linear Diblock Copolymers, *Phys Rev Lett*, 2017,**118**, 207901.
[8] Gillard T M, Lee S and Bates F S, Dodecagonal quasicrystalline order in a diblock copolymer melt,*Proc. Natl. Acad. Sci*, 2016, **113**, 5167-5172.
[9]Xiao C, Fujita N, Miyasaka K, Sakamoto Y, Terasaki O, Dodecagonal tiling inmesoporoussilica,*Nature*, 2012,**487**(7407), 349–353.
[10]Fischer S, Exner A, Zielske K, Perlich J, Deloudi S, Steuer W, Linder P and Foestor S, Colloidal quasicrystals with 12-fold and 18-fold diffraction symmetry, *Proc.Natl. Acad. Sci.*,2011,108 1810-1814.
[11]Stephen Z. D. Cheng,Giant Surfactants based on Precisely Functionalized POSS Nano-atoms: Tuning from Crystals to Frank-Kasper Phases and Quasicrystals, *First Annual Symposium on Frontiers of Soft Matter Science and Engineering*, 2015, Beijing/2016, Bulletin of the American Physical Society.
[12] Yue K, Huang M J, R Marson, He J L, Huang J H, Zhou Z, Liu C, Yan X S, Wu K, Wang J, Guo Z H, Liu H, Zhang W, Ni P H,Wesdemiotis C, Zhang W B, Glotzer S C and Cheng S Z D, Geometry induced sequence of nanoscale Frank-Kasper and quasicrystalmesophases in giant surfactants, *Proc. Natl. Acad. Sci.,* 2016,**113**, 1392-1400.
[13]P. G. de Gennes, Soft matter,.*Rev.Mod. Phys,*1992,**64**(3), 645-648.
[14]UngarGoran and ZengXiangbing,Frank–Kasper, quasicrystalline and related phases in liquidcrystals, *Soft Matter*, 2005,**1**(2),95–106

[15] Huang M, Yue K, Wang J, Hsu C-H, Wang L and Cheng S. Z. D. Frank-Kasper and related quasicrystal spherical phases inmacromolecules, *Science ChinaChemistry*, 2018, **61**, 33-45.

[16] Kim S A, JeongKJ, YethirajA, andMahanthappa M K, Low-symmetry sphere packings of simple surfactant micelles induced by ionic sphericity, *Proc.Natl. Acad. Sci.*, 2017, **114**, 4072-4077.

[17] Carlos M. Baez-Cotto and Mahesh K. Mahanthappa, Micellar Mimicry of Intermetallic C14 and C15 Laves Phases by Aqueous Lyotropic Self-Assembly, *ACS Nano*, 2018, **12**(4), 3226-323.

[18] Kim H, Song Z, Leal C, Super-swelled lyotropic single crystals, *Proc.Natl Acad. Sci.*, 2017, **114**, 10834-10839.

[19] Lifshitz R and DiamantH, Soft quasicrystals - Why are they stable? *Phil. Mag.*, 2007, **87**, 3021-3030.

[20] Barkan K, Diamant H and Lifshitz R, Stability of quasicrystals composed of soft isotropic particles *Phys. Rev. B*, 2011, **83**, 172201.

[21] K. Jiang, J. Tong, P. Zhang, and A.-C. Shi, Stabilityoftwo-dimensional soft quasicrystals in systems with twolengthscales, *Phys. Rev. E*, 2015, **92**, 042159.

[22] S. Savitz, M. Babadi, and R. Lifshitz, Multiple-scalestructures: from Faraday waves to soft-matter quasicrystals, *International Union of Crystallography Journal*, 2018, **5**(3), 247-268.

[23] K. Jiang and W. Si, Stability of three-dimensional icosahedralquasicrystals in multi-component systems," arXivpreprint arXiv:1903.07859, 19 Mar 2019.

[24] D.J. Ratliff, A.J. Archer, P. Subramanian, A.M. Rucklidge, Which wavenumbers determine the thermodynamic stability of soft matter quasicrystals?, arXiv:1907.05805, 10 Jul 2019.

[25] Lifshitz R and Petrich D M, Theoretical Model for Faraday Waves with Multiple-Frequency Forcing, arXiv:cond-mat/9704060v2 [cond-mat.soft], 29 Jan 1998.

[26] Fan T Y, Equation Systems of Generalized Hydrodynamics forSoft-Matter Quasicrystals, *Appl. Math. Mech.*, 2016, **37**(4), 331-347 (in Chinese), arXiv: 1908.06425[cond-mat.soft], 18 Aug 2019.

[27] Fan T Y, Generalized Dynamics for Second Kind of Soft-Matter Quasicrystals, *Appl. Math.Mech.*, 2017, **38**(2), 189-199(in Chinese), arXiv:1908.06430[cond-mat.soft], 18 Aug 2019.

[28] Fan T Y and Tang Z Y, Three-Dimensional Generalized Dynamics of Soft-Matter Quasicrystals, *Appl. Math.Mech.*, 2017, **38**(11), 1195-1207 (in Chinese), arXiv:1908.06547[cond-mat.soft], 19 Aug 2019.

[29] Fan T Y, Generalized Dynamics of Soft-Matter Quasicrystals---Mathematical Models and Solutions, 2017, Beijing, Beijing Institute of Technology Press, / Heidelberg, Springer-Verlag.

[30] Lubensky T C, Ramaswamy S and Toner J, Hydrodynamics of icosahedral quasicrystals, *Phys. Rev. B*, 1985, **32**, 7444-7452.

[31] Cowley R A, Acoustic phonon instabilities and structural phase transitions, *Phys. Rev. B*, 1976, **13**, 4877-4885.

[32] Fan T Y, Mathematical Theory of Elasticity of Quasicrystals and Its Applications, 2010 1st edition, 2016 2nd edition, Beijing, Science Press / Heidelberg, Springer-Verlag.

[33]Tang Z Y and Fan T Y, The stability of the second kind of soft matter quasicrystals, 2019, to be submitted.